\documentclass[12pt,onecolumn,floatfix,amsmath,nofootinbib,amssymb,floatfix]{revtex4}
\usepackage{}
\usepackage{graphicx,color,dcolumn,booktabs,bm}
\usepackage{longtable,lscape}
\usepackage{txfonts}
\usepackage{float}
\usepackage{slashed}
\usepackage{makecell}
\usepackage{tabularx}
\usepackage{array}
\usepackage{ragged2e}
\newcolumntype{P}[1]{>{\centering\hspace{0pt}}p{#1}}
\newcolumntype{Z}{>{\centering\let\newline\\\arraybackslash\hspace{0pt}}x}

\usepackage{booktabs}
\usepackage{overpic}
\usepackage{amssymb}
\usepackage{indentfirst}
\usepackage{feynmf}   
\usepackage{slashed}  
\usepackage{cases}
\usepackage{color}
\usepackage{multirow}
\usepackage{epstopdf}
\usepackage{epsfig}
\usepackage{longtable}
\usepackage[colorlinks,
            citecolor=blue,
            anchorcolor=red,
            menucolor=red,
            linkcolor=red,
            filecolor=red,
            runcolor=red,
            urlcolor=blue,
            frenchlinks=red]{hyperref}
\usepackage{hyperref}
\usepackage{multirow}

\graphicspath{{Figures/}} %
\allowdisplaybreaks
\usepackage{indentfirst}

\begin{document}
\title{Study of form factors and branching ratios for $D\rightarrow S, Al\bar{\nu_{l}}$ with light-cone sum rules}
\author{Qi Huang$^{1,2}$}
\email{qihuang1193572279@163.com}
\author{Yan-Jun Sun$^{1,2}$}
\email{sunyanjun@nwnu.edu.cn}
\author{Di Gao$^{1,2}$}
\email{d-gao@foxmail.com}
\author{Guo-Hua Zhao$^{1,2}$}
\email{guohuazhao0916@163.com}
\author{Bin Wang$^{1,2}$}
\email{wangbin200006@outlook.com}
\author{Wei Hong$^{1,2}$}
\email{hongwei17809213575@163.com}
\affiliation{ $^1$ College of Physics and Electronic Engineering, Northwest Normal University, Lanzhou 730070, China\\
$^2$ Lanzhou Center for Theoretical Physics, Lanzhou University Lanzhou 730070, China}
\begin{abstract}
We systematically study the semileptonic decay process of $ D\rightarrow S, A l\bar{\nu_{l}}(l=e,\mu)$ by light-cone sum rules (LCSR) with chiral currents, calculate the form factors containing only the contribution of the leading twist light-cone distribution amplitudes (LCDAs). For scalar mesons $a_{0}(980)$ and  $a_{0}(1450)$, we take them as $q\bar{q}$ states. For axial-vector meson, we study $a_{1}(1260)( 1^{3}p^{1})$ and $b_{1}(1235)( 1^{1}p^{1})$. Based on the results of these form factors, we further present the branching ratios of these semileptonic decay processes. The numerical results for $ D\rightarrow a_{0}(980), b_{1}(1235)l\bar{\nu_{l}} $ are in good agreement with experiments and that for $ D\rightarrow a_{0}(1450), a_{1}(1260)l\bar{\nu_{l}}$ processes are expected to be tested experimentally in the future.
\end{abstract}
\maketitle
\section{Introduction}
\setlength{\parindent}{1em}
The study of the $D$ meson semileptonic decay process can not only be used to extract the CKM (Cabibbo-Kobayashi-Maskawa) matrix elements, but also play an important
role in determining weak interaction between quarks in the standard model (SM). More specifically, the $D$ meson semileptonic decay process has a relatively simple decay mechanism and final state interactions, making it an ideal occasion for studying light scalar meson structures. To figure it out, a series of experiments have been carried out by BES \uppercase\expandafter{\romannumeral3} \cite{Ablikim:2018ffp,Ablikim:2020agq,Ablikim:2017lks,Ablikim:2018evp}, CLEOc \cite{Besson:2009uv,CLEO:2011ab}, BarBar \cite{Lees:2014ihu}, Belle \cite{Widhalm:2006wz}, and so on. 

For $D\rightarrow P, V l\bar{\nu_{l}}$ ($P, V$ are pseudoscalar and vector mesons respectively), a lot of research work on experiments \cite{Ablikim:2017lks,Ablikim:2018evp,Besson:2009uv,CLEO:2011ab,Lees:2014ihu,Widhalm:2006wz} and theories \cite{Palmer:2013yia,Li:2019phv,Bernard:1991bz,Wang:2002zba,Wu:2006rd,Li:2012gr,Khodjamirian:2009ys,Ball:2006yd,Khodjamirian:2000ds,Ball:1993tp,Faustov:2019mqr,Soni:2018adu,Fu:2018yin,Chang:2019mmh,Fajfer:2005ug,Huang:2008sn} have been carried out, but less for $D\rightarrow S, A l \bar{\nu_{l}}$. From the theoretical point of view, the decay channels of $D\rightarrow P, V l \bar{\nu_{l}}$ can be studied directly because the final state mesons are classical quark-antiquark states, but for the scalar mesons below 1 GeV,  there have been many controversy about their internal structures. For example, QCD sum rules (QSR) has investigated the possibility to distinguish the two-quark and tetra-quark picture for light scalar meson \cite{Wang:2009azc} and LCSR has studied them as quark-antiquark states\cite{Cheng:2017fkw}. Lattice QCD has studied them as tetra-quark states \cite{Alexandrou:2017itd}, and the MIT bag model has studied them with a diquark-diantiquark picture \cite{Jaffe:1976ig}. Throughout this paper, we take scalar meson $a_{0}(980)$ and $a_{0}(1450) $ as quark-antiquark state.
For axial-vector meson, there are two different nonets of $J^{P} = 1^{+}$ in the quark model as the orbital excitations of the $q\bar{q}$ system. In terms of the spectroscopic notation $^{2s+1}L_{J}$, p-wave axial-vector meson can be divided into $1^{3}p^{1}$ ($C+$) and $1^{1}p^{1}$ ($C-$) depending on the $C$-quantum number. Experimentally, the $J^{PC} = 1^{++}$ nonet consists of
$a_{1}(1260)$, $f_{1}(1285)$, $f_{1}(1420)$ and $K_{1A}$, while the $J^{PC} = 1^{+-}$ nonet has $b_{1}(1235)$, $h_{1}(1170)$, $h_{1}(1380)$
and $K_{1B}$.
For the present work, we study two mesons $a_{1}(1260)( 1^{3}p^{1})$  and  $b_{1}(1235)( 1^{1}p^{1})$ with simple internal structure relative to  other axial-vector mesons, which cannot have mixing due to the opposite $C$-parities.

The form factors which are important parameters for theoretical studies of the $D$ meson semileptonic decay process, can be calculated by various methods that are powerful in a certain region of the transfer momentum square $q^{2}$. For example, the low energy effective theory (LEET) can be used in the region $q^{2}\rightarrow0$ \cite{Palmer:2013yia}; lattice QCD (LQCD) can be used in the region of large momentum transfer $q^{2}\rightarrow\infty$ \cite{Li:2019phv,Bernard:1991bz}; the heavy quark effective theory (HQET) and the heavy-light chiral perturbation theory can be used in the region $q^{2}=\infty$ \cite{Wang:2002zba,Wu:2006rd}.

In this paper, we study the form factors of semileptonic decay process $D\rightarrow S, A l \bar{\nu_{l}}$ by using the LCSR method with the chiral currents, in the region $0 \leqslant q^{2} \leqslant (m_{D}-m_{M})^{2}$ ($M$ are scale meson or axial-vector meson) .
Experimentally, BES \uppercase\expandafter{\romannumeral3} \cite{Ablikim:2018ffp,Ablikim:2020agq} provide an ideal place to study $D\rightarrow S, A l \bar{\nu_{l}}$. 
 To compare with it, we also calculate the branching ratios.

The paper is organized as follows: In Sec. II,
the form factors and branching ratios of the semileptonic decays for $D\rightarrow S, A l \bar{\nu_{l}}$ are derived, and some simple relations between form factors are also obtained. In Sec. III, we present our numerical analysis on values of the form factors and branching ratios of the semileptonic decays, and a comparison is also made between our results and the predictions of other methods and experiments.

\section{Form Factors}

In the semileptonic decays, the hadron currents, which represent the strong interaction between quarks, can be parameterized into form factors.
The hadronic matrix elements for $D\rightarrow S$ can be parameterized by form factors as:
\begin{align}
\langle S(p)|\bar{q}_{2}\gamma_{\mu}\gamma_{5}c|D(p+q)\rangle&=-2if_{+}^{D\rightarrow S}(q^{2})p_{\mu}-i[f_{+}^{D\rightarrow S}(q^{2})+f_{-}^{D\rightarrow  S}(q^{2})]q_{\mu},\tag{1}   \label{1}
\end{align}
\begin{align}
\langle S(p)|\bar{q}_{2}\sigma_{\mu\nu}(1+\gamma_{5})q^{\nu}c|D(p+q)\rangle&=[2p_{\mu}q^{2}-2q_{\mu}(p\cdot q)]\frac{-f_{T}^{D\rightarrow S(q^2)}}{m_{D}+m_{S}},\tag{2}  \label{2}
\end{align}
where $f_{+}^{D\rightarrow S}(q^{2})$ and $f_{-}^{D\rightarrow S}(q^{2})$ are the transition form factors of the $D\rightarrow S$, and $f_{T}^{D\rightarrow S}(q^{2})$ is the penguin form factor. The hadronic matrix elements for $D\rightarrow A$ can be parameterized by form factors as:
\begin{align}
\langle A(p)|\bar{q}_{2}\gamma_{\mu}(1+\gamma_{5})c|D(p+q)\rangle=&\frac{2iA^{D\rightarrow A}(q^2)}{m_{D}-m_{A}}\epsilon_{\mu\nu\alpha\beta}\varepsilon^{*\nu}(q+p)^{\alpha}p^{\beta}\notag\\
&-(m_{D}-m_{A})V_{1}^{D\rightarrow A}(q^2)\varepsilon^*_{\mu}\notag\\
&+\frac{V_{2}^{D\rightarrow A}(q^2)}{m_{D}-m_{A}}(\varepsilon^*\cdot{q})(2p+q)_{\mu}\notag\\
&+\frac{2(\varepsilon^*\cdot{q})m_{A}}{q^{2}}q_{\mu}[V_{3}^{D\rightarrow A}(q^2)-V_{0}^{D\rightarrow A}(q^2)]\tag{3} \label{3}
\end{align}
and
\begin{align}
\langle A(p)|\bar{q}_{2}\sigma_{\mu\nu}(1+\gamma_{5})q^{\nu}c|D(p+q)\rangle=&-2\epsilon_{\mu\nu\rho\sigma}\varepsilon^{*\nu}q^{\rho}p^{\sigma}T_{1}^{D\rightarrow A}(q^2),\notag\\
&+i(\varepsilon^*\cdot {q})p_{\mu}[T_{2}^{D\rightarrow A}(q^2)+(\frac{q^2}{m_{D}^2-m_{V}^2}-1)T_{3}^{D\rightarrow A}(q^2)]\notag\\
&+2i(\varepsilon^*\cdot {q})p_{\mu}(T_{2}^{D\rightarrow A}(q^2)+\frac{q^2}{m_{D}^2-m_{V}^2}T_{3}^{D\rightarrow A}(q^2))\notag\\
&-i\varepsilon^*_{\mu}(m_{D}^2-m_{A}^2)T_{2}^{D\rightarrow A}(q^2),\tag{4}  \label{4}
\end{align}
where
\begin{align}
V_{3}^{D\rightarrow A}(q^2)&=\frac{m_{D}-m_{A}}{2m_{A}}V_{1}^{D\rightarrow A}(q^2)-\frac{m_{D}+m_{A}}{2m_{A}}V_{2}^{D\rightarrow A}(q^2),\notag\\
V_{0}^{D\rightarrow A}(q^2)&=\frac{m_{D}-m_{A}}{2m_{A}}V_{1}^{D\rightarrow A}(q^2)-\frac{m_{D}+m_{A}}{2m_{A}}V_{2}^{D\rightarrow A}(q^2)+\frac{q^{2}}{2m_{A}(m_{D}-m_{A})}V_{2}^{D\rightarrow A}(q^2),\tag{5}  \label{5}
\end{align}
and $\varepsilon$ is polarization vector for axial-vector meson.

To calculate the form factors, we construct the two-point correlation function as
\begin{align}
\Pi_{1\mu}(p,q)&=i\int d^{4}xe^{iqx}\langle M(p)|T\{J_{1\mu}(x),J_{1}(0)\}|0\rangle,\tag{6}  \label{6}
\end{align}
\begin{align}
\Pi_{2\mu}(p,q)&=i\int d^{4}xe^{iqx}\langle M(p)|T\{J_{2\mu}(x),J_{1}(0)\}|0\rangle,\tag{7}     \label{7}
\end{align}
The transfer momentum $q$ is defined as $q=p_{D}-p$, where $p_{D}$ and $p$ are the four-momentum of the initial and final meson states. $J_{i\mu}$ and $J_{i}$ $(i=1,2)$ are chiral currents, whose explicit forms are shown in Table I, depending on the decay processes.
\begin{table}[H]
	\centering
\caption{Chiral currents for different decay processes.}
	\begin{tabular}{p{4cm}<{\centering}p{5cm}<{\centering}p{5cm}<{\centering}}
		\hline
        Mode &  $J_{i\mu}(x)$&  $J_{i}(0)$  \\
        \hline

        \multirow{2}*{$ D\rightarrow S $} & $J_{1\mu}(x)=\bar{q}_{2}(x)\gamma_{\mu}(1-\gamma_{5})c(x)$ & $J_{1}(0)=\bar{c}(0)i(1-\gamma_{5})q_{1}(0)$  \\
                                          & $J_{2\mu}(x)=\bar{q}_{2}(x)\sigma_{\mu\nu}(1+\gamma_{5})c(x)$ & $J_{2}(0)=\bar{c}(0)i(1-\gamma_{5})q_{1}(0)$  \\
	
		\cline{1-3}
		
		\multirow{2}*{$ D\rightarrow A $} & $J_{1\mu}(x)=\bar{q}_{2}(x)\gamma_{\mu}(1-\gamma_{5})c(x)$ & $J_{1}(0)=\bar{c}(0)i(1+\gamma_{5})q_{1}(0)$  \\
                                          & $J_{2\mu}(x)=\bar{q}_{2}(x)\sigma_{\mu\nu}(1+\gamma_{5})c(x)$ & $J_{2}(0)=\bar{c}(0)i(1+\gamma_{5})q_{1}(0)$  \\

        \hline
	\end{tabular}
\end{table}

In general, the correlation function should be described by two ways:

(\textbf{1}) Inserting a complete set of intermediate hadronic states in the middle of two currents in the correlation function to obtain the phenomenological side;

(\textbf{2}) Operator product expanding (OPE) for correlation functions to obtain the theoretical side.

\textbf{On the phenomenological side}

The hadronic representations of correlators are achieved by inserting between two currents a complete set of resonance states with the same quantum numbers as the operator $J_{1}(0)$. On the desired pole contributions due to the lowest pseudoscalar $D$ meson are insolated and we obtain the hadronic repredentations 
\begin{align}
\Pi_{1\mu}(p,q)&=\frac{\langle M(p)|\bar{q}_{2}\gamma_{\mu}(1\pm \gamma_{5}) c|D\rangle \langle D|\bar{c}i\gamma_{5}q_{1}|0\rangle}{m^{2}_{D}-(p+q)^{2}}+higher \quad states,\tag{8}
\end{align}
\begin{align}
\Pi_{2\mu}(p,q)&=\frac{\langle M(p)|\bar{q}_{2}\sigma_{\mu\nu}(1\pm \gamma_{5})q^{\nu}c|D\rangle \langle D|\bar{c}i\gamma_{5}q_{1}|0\rangle}{m^{2}_{D}-(p+q)^{2}}+higher \quad states.\tag{9}
\end{align}
It should be stressed that the correlation functions receive contributions from the scalar resonances, in addition to the higher pseudoscalar ones, and the ground-state scalar meson is a bit lighter than the pseudoscalar resonance lying in the first excited state.  The matrix element, $\langle M(p)|\bar{q}_{2}\gamma_{\mu}(1\pm \gamma_{5}) c|D\rangle $ and $\langle M(p)|\bar{q}_{2}\sigma_{\mu\nu}(1\pm \gamma_{5})q^{\nu}c|D\rangle $, can be parameterized by the form factors as in Eqs. (\ref{1})-(\ref{5}).
The second element is expressed via $D$ meson decay constant as:
\begin{align}
\langle D|\bar{c}i\gamma_{5}q_{1}|0\rangle&=\frac{m^{2}_{D}f_{D}}{m_{c}+m_{q_{1}}}.\tag{10}
\end{align}
Substituting the matrix elements Eqs. (\ref{1})-(\ref{5}) into Eqs. (\ref{6}) and  (\ref{7}), we obtain the phenomenological part of correlation function in terms of the form factors and Lorentz structures. The correlation functions of $D\rightarrow S$ are expressed as
\begin{align}
\Pi_{1\mu}(p,q)&=-\frac{2if_{+}^{D\rightarrow S}(q^{2})p_{\mu}+i[f_{+}^{D\rightarrow S}(q^{2})+f_{-}^{D\rightarrow S}(q^{2})]q_{\mu}}{m_{D}^{2}-(p+q)^{2}}\frac{m^{2}_{D}f_{D}}{m_{c}+m_{q_{1}}}+\int_{s_{0}}^{\infty}\frac{\rho(s)ds}{s-(p+q)^{2}},\tag{11}
\end{align}
\begin{align}
\Pi_{2\mu}(p,q)&=-\frac{[q^{2}2p_{\mu}-2q_{\mu}(p\cdot q)]}{m_{D}^{2}-(p+q)^{2}}\frac{if_{T}^{D\rightarrow S}}{m_{D}+m_{P}}\frac{m^{2}_{D}f_{D}}{m_{c}+m_{q_{1}}}+\int_{s_{0}}^{\infty}\frac{\rho(s)ds}{s-(p+q)^{2}},\tag{12}
\end{align}
and the correlation functions of $D\rightarrow A$ as
\begin{align}
\Pi_{1\mu}(p,q)=&\{\frac{2iA^{D\rightarrow A}(q^2)}{m_{D}-m_{A}}\epsilon_{\mu\nu\alpha\beta}\varepsilon^{*\nu}(q+p)^{\alpha}p^{\beta}-(m_{D}-m_{A})V_{1}^{D\rightarrow A}(q^2)\varepsilon^*_{\mu}+\frac{V_{2}^{D\rightarrow A}(q^2)}{m_{D}-m_{A}}(\varepsilon^*\cdot{q})(2p+q)_{\mu}\notag\\
&+\frac{2(\varepsilon^*\cdot{q})m_{A}}{q^{2}}q_{\mu}[V_{3}^{D\rightarrow A}(q^2)-V_{0}^{D\rightarrow A}(q^2)]\}\frac{m^{2}_{D}f_{D}}{m_{c}+m_{q_{1}}}+\int_{s_{0}}^{\infty}\frac{\rho(s)ds}{s-(p+q)^{2}},\tag{13}
\end{align}
\begin{align}
\Pi_{2\mu}(p,q)=&\{-2\epsilon_{\mu\nu\rho\sigma}\varepsilon^{*\nu}q^{\rho}p^{\sigma}T_{1}^{D\rightarrow A}(q^2)+i(\varepsilon^*\cdot{q})p_{\mu}[T_{2}^{D\rightarrow A}(q^2)
+(\frac{q^2}{m_{D}^2-m_{A}^2}-1)T_{3}^{D\rightarrow A}(q^2)]\}\notag\\
&+2i(\varepsilon^*\cdot{q})p_{\mu}[T_{2}^{D\rightarrow A}(q^2)+\frac{q^2}{m_{D}^2-m_{A}^2}T_{3}^{D\rightarrow A}(q^2)]-i\varepsilon^*_{\mu}(m_{D}^2-m_{A}^2)T_{2}^{D\rightarrow A}(q^2)\}\notag\\
&\frac{1}{m_{D}^{2}-(p+q)^{2}}\frac{m^{2}_{D}f_{D}}{m_{c}+m_{q_{1}}}+\int_{s_{0}}^{\infty}\frac{\rho(s)ds}{s-(p+q)^{2}},\tag{14}
\end{align}
 where $s_{0}$ near the squared mass of the lowest scalar $D$ meson and $\rho(s)$ is the spectral density of the higher excited states and continuum. Since the contributions of the higher excited and continuum spectrum on the phenomenological side cannot be calculated exactly, we apply quark-hadron duality to express the integral of the higher excited and continuum spectrum\cite{Colangelo:2000dp}:
\begin{align}
\int_{s_{0}}^{\infty}ds\frac{\rho(s)}{s-(p+q)^{2}}\simeq \int_{s_{0}}^{\infty}ds\frac{1}{\pi}\frac{Im\Pi_{\mu}^{pert}(p,q)}{s-(p+q)^{2}}.\tag{15}
\end{align}

\textbf{On the theoretical side}

The calculation of the correlation function in the region of large space-like momentum is based on the expansion of the T-product of the quark currents near the light-cone $x^{2}\approx 0$, due to sufficiently large momentum transfer. By contracting $c$ and $\bar{c}$ quark fields, we obtain
\begin{align}
\Pi_{1\mu}(p,q)&=i\int d^{4}xe^{iqx}\langle M(p)|\{\bar{q}_{2}(x)\gamma_{\mu}(1\pm\gamma_{5})S^{c}(x,0)i(1\pm\gamma_{5})q_{1}(0)\}|0\rangle,\tag{16}
\end{align}
\begin{align}
\Pi_{2\mu}(p,q)&=i\int d^{4}xe^{iqx}\langle M(p)|\{\bar{q}_{2}(x)\sigma_{\mu\nu}(1\pm\gamma_{5})q^{\nu}S^{c}(x,0)i(1\pm\gamma_{5})q_{1}(0)\}|0\rangle,\tag{17}
\end{align}
where $S^{c}(x,0)$ is the propagator of the free-$c$ quark as
\begin{align}
S^{c}(x,0)&=-i\int\frac{d^{4}k}{(2\pi)^{4}}e^{-ik\cdot x}\frac{\slashed{k}+m_{c}}{m^{2}_{c}-k^{2}}.\tag{18}
\end{align}
After rearrangement of the quantum fields and matrices, the correlation function, is turned into a form including a matrix trace and a matrix element of non-local operators between $M$ meson state and vacuum state,
\begin{align}
\Pi_{1\mu}(p,q)=&i\int\frac{d^{4}xd^{4}k}{(2\pi )^{4}}\frac{e^{i(q-k)x}}{m_{c}^{2}-k^{2}}Tr{[\gamma_{\mu}(1\pm\gamma_{5})(\slashed{k}+m_{c})(1\pm\gamma_{5})]_{\delta\alpha}\langle M(p)|\bar{q}_{2\delta}(x)q_{1\alpha}(0)}|0\rangle,\tag{19}  \label{19}
\end{align}
\begin{align}
\Pi_{2\mu}(p,q)=&i\int\frac{d^{4}xd^{4}k}{(2\pi )^{4}}\frac{e^{i(q-k)x}}{m_{c}^{2}-k^{2}}Tr{[\sigma_{\mu\nu}(1\pm\gamma_{5})q^{\nu}(\slashed{k}+m_{c})(1\pm\gamma_{5})]_{\delta\alpha}\langle M(p)|\bar{q}_{2\delta}(x)q_{1\alpha}(0)}|0\rangle.\tag{20}   \label{20}
\end{align}

In the LCSR method, the non-vanishing matrix elements are defined in terms of LCDAs. Generally, the matrix element $\langle S(p)|\bar{q}_{\delta}(x)q_{\alpha}(0)|0\rangle$ is written as \cite{Cheng:2005nb}
\begin{align}
\langle S(p)|\bar{q}_{\delta}(x)q_{\alpha}(0)|0\rangle&=\frac{1}{4}\int_{0}^{1}due^{iup\cdot x}\{\slashed{p}\phi_{S}(u)+m_{S}(\phi^{s}_{S}(u)-\sigma_{\mu\nu}p^{\mu}x^{\nu}\frac{\phi^{\sigma}_{s}(u)}{6})\}_{\delta\alpha},\tag{21}  \label{21}
\end{align}
where $\phi_{S}(u)$ is the twist-2 LCDA, $\phi^{s}_{S}(u)$ and $\phi^{\sigma}_{s}(u)$ are twist-3 LCDAs of scalar meson. Similarly, the matrix element $\langle A(p)|\bar{q}_{\delta}(x)q_{\alpha}(0)|0\rangle$ is expressed as \cite{Yang:2008xw}
\begin{align}
\langle A(p,\varepsilon^{*})|\bar{q}_{\delta}(x)q_{\alpha}(0)|0\rangle=&-\frac{1}{4}\int^{1}_{0}due^{i(up\cdot{x})}\times\{f_{A}m_{A}[\slashed{p}\gamma_{5}
\frac{\varepsilon^{*} x}{px}(\phi_{\parallel}(u)+\frac{m_{A}^2x^2}{16}A_{\parallel}^2)\notag\\
&+(\slashed{\varepsilon}^{*}-\slashed{p}{\varepsilon^{*} x}{px})\gamma_{5}g^{(a)}_{\bot}(u)-\slashed{x}\gamma_{5}\frac{\varepsilon^{*}x}{2(px)^2}m_{A}^2\bar{g}_{3}(u)+\epsilon_{\mu\nu\rho\sigma}\varepsilon^{*\nu}
p^{\rho}x^{\sigma}\gamma^{\mu}
\frac{g^{(v)}_{\bot}(u)}{4}]\notag\\
&+f_{\bot}^{A}[\frac{1}{2}(\slashed{p}\slashed{\varepsilon}^{*}-\slashed{\varepsilon}^{*}\slashed{p})\gamma_{5}(\phi_{\bot}(u)+\frac{m_{A}^2x^2}{16}A_{\bot}^2)
-\frac{1}{2}(\slashed{p}\slashed{x}-\slashed{x}\slashed{p})\gamma_{5}\frac{\varepsilon^{*}x}{2(px)^2}m_{A}^2\bar{h}_{s}(u)\notag\\
&-\frac{1}{4}(\slashed{\varepsilon}^{*}\slashed{x}-\slashed{x}\slashed{\varepsilon}^{*})\gamma_{5}\frac{m_{A}^2}{px}m_{A}^2\bar{h}_{3}(u)+i(\varepsilon^{*}x)
m_{A}^2\gamma_{5}\frac{h_{\parallel}^{(p)}(u)}{2}]\big\}_{\delta\alpha},\tag{22}
\end{align}
where $\phi_{\parallel}(u)$ and $\phi_{\bot}(u)$ are twist-2 LCDAs, $g^{(a)}_{\bot}$, $g^{(v)}_{\bot}$, $h_{\parallel}^{(p)}$ and $h_{\parallel}^{(t)}$ are twist-3 LCDAs, and $g_{3}$ and $h_{3}$ are twist-4 LCDAs of axial-vector meson,
\begin{align}
\bar{g}_{3}(u)&=g_{3}(u)+\phi_{\parallel}(u)-2g_{\bot}^{a}(u),\notag\\
\bar{h}_{\parallel}^{t}&=h_{\parallel}^{h}-\frac{1}{2}\phi_{\bot}(u)-\frac{1}{2}h_{3}(u),\notag\\
\bar{h}_{3}(u)&=h_{3}(u)-\phi_{\bot}(u). \tag{23}
\end{align}

For $D\rightarrow S$, substituting Eq. (\ref{21}) into correlation functions (\ref{19}) and (\ref{20}), and making a trace , we obtain the correlator on the theoretical side as
\begin{align}
\Pi_{1\mu}&=2im_{c}p_{\mu}\int^{1}_{0}du\frac{\phi_{S}(u)}{m^{2}_{c}-(q+up)^{2}},\tag{24}
\end{align}
\begin{align}
\Pi_{2\mu}&=-[2p_{\mu}q^{2}-2q_{\mu}(p\cdot q)]\int^{1}_{0}du\frac{\phi_{S}(u)}{m^{2}_{c}-(q+up)^{2}}.\tag{25}
\end{align}
Different from the process $D\rightarrow S$, for $D\rightarrow A$ process, axial-vector meson LCDAs are usually expressed as longitudinal and transverse projection operator \cite{Yang:2008xw}. The transverse projection of the correlation functions are
\begin{align}
\Pi_{1\mu}(p,q)=&i\int\frac{d^{4}xd^{4}k}{(2\pi )^{4}}\frac{e^{i(q-k)x}}{m_{c}^{2}-k^{2}}Tr{[\gamma_{\mu}(1-\gamma_{5})(\slashed{k}+m_{c})(1+\gamma_{5})]}_{\delta\alpha}M^{A}_{\bot\delta\alpha},\tag{26}  \label{26}
\end{align}
\begin{align}
\Pi_{2\mu}(p,q)=&i\int\frac{d^{4}xd^{4}k}{(2\pi )^{4}}\frac{e^{i(q-k)x}}{m_{c}^{2}-k^{2}}Tr{[\sigma_{\mu\nu}(1+\gamma_{5})q^{\nu}(\slashed{k}+m_{c})(1+\gamma_{5})]}_{\delta\alpha}
M^{A}_{\bot\delta\alpha},\tag{27} \label{27}
\end{align}
where $M^{A}_{\bot\delta\alpha}$ is transverse projection operator \cite{Yang:2008xw},
\begin{align}
M^{A}_{\bot\delta\alpha}=&i\frac{f_{\bot}}{4}E\{\slashed{\varepsilon}^*_{\bot}\slashed{n}_{-}\gamma_{5}\phi_{\bot}(u)-\frac{f_{V}}{f_{A}^{\bot}}\frac{m_{A}}{E}
[\slashed{\varepsilon}^{*}_{\bot}\gamma_{5}g^{(a)}_{\bot}(u)\notag\\
&-E\int^{u}_{0}dv(\phi_{\parallel}(v)-g^{(a)}_{\bot}(v))\slashed{n}_{-}\gamma_{5}\varepsilon^{*}_{\bot\mu}\frac{\partial}{\partial{k_{\bot\mu}}}\notag\\
&+i\epsilon_{\mu\nu\rho\sigma}\gamma_{\mu}\varepsilon^{*\nu}n_{-}^{\rho}n_{+}^{\sigma}\frac{g^{(v)'}_{\bot}(u)}{8}-E\frac{g^{(v)}_{\bot}(u)}{4}
\frac{\partial}{\partial{k_{\bot\sigma}}}]\}\mid _{k=up}+\mathcal {O}(\frac{m_{A}^{2}}{E^{2}}).\tag{28}  \label{28}
\end{align}
Substituting Eq. (\ref{28}) into correlation functions (\ref{26}) and (\ref{27}), and performing the trace operation for Eq. (\ref{26}) and Eq. (\ref{27}), respectively, we obtain the correlator of $D\rightarrow A$ on the theoretical side as 
\begin{align}
\Pi_{1\mu}=&-2f^{A}_{\bot}\int^{1}_{0}du[-p\cdot{(q+up)}\varepsilon^{*}_{\bot\mu}-(\varepsilon^*_{\bot}\cdot{q})p_{\mu}
-i\epsilon_{\mu\nu\beta\tau}\varepsilon^{*\beta}_{\bot}q^{\mu}p^{\tau}]\frac{\phi_{\bot}(u)}{m^{2}_{c}-(q+up)^{2}},\tag{29}
\end{align}
\begin{align}
\Pi_{2\mu}=&2f^{A}_{\bot}m_{c}\int^{1}_{0}du[i(q\cdot{p})\varepsilon^{*}_{\bot\mu}-i(\varepsilon^{*}_{\bot}\cdot{q})p_{\mu}
-\epsilon_{\mu\nu\rho\tau}\varepsilon^{*\rho}_{\bot}q^{\nu}p^{\tau}]\frac{\phi_{\bot}(u)}{m^{2}_{c}-(q+up)^{2}}. \tag{30}
\end{align}

 Base on the conformal symmetry hidden in the QCD Lagrangian, $\phi_{S}(u)$ and $\phi_{\bot}(u)$ can be expanded in a series of
 Gegenbauer polynomials $C_{m}^{3/2}$ with increasing conformal spin as \cite{Braun:2003rp}
\begin{align}
\phi_{S}(u,\mu)=\bar{f_{S}}(\mu)6u(1-u)[B_{0}(\mu)+\sum_{m=1}B_{m}(\mu)C_{m}^{3/2}(2u-1)]\tag{31}
\end{align}
for scalar mesons, where $B_{m}(\mu)$ is Gegenbauer coefficient,
\begin{align}
 B_{m}(\mu)=\frac{1}{\bar{f_{S}}}\frac{2(2m+1)}{3(m+1)(m+2)}\int_{0}^{1}C_{m}^{3/2}(2u-1)\phi_{S}(u,\mu)du\tag{32}
\end{align}
and
\begin{align}
\bar{f_{S}}(\mu)&=\bar{f_{S}}(\mu_{0})(\frac{\alpha_{s}(\mu_{0})}{\alpha_{s}(\mu)})^{4/b}.\tag{33}
\end{align}
In the SU(3) limit, $B_{0}=0$, and the twist-2 LCDAs of all scalar mesons are antisymmetric in the $u\rightarrow1-u$ transformation, thus, only odd Gegenbauer coefficients ($B_{1}(\mu), B_{3}(\mu)$) are considered in the following discussion.
For axial-vector meson, the twist-2 LCDAs are \cite{Yang:2007zt}
\begin{align}
\phi_{\bot}^{A}(u)=6u(1-u)[a_{0}^{\bot}+3a_{1}^{\bot}\xiup +a_{2}^{\bot}\frac{3}{2}(5\xiup ^{2}-1)]\tag{34}
\end{align}
for $1^{3}P_{1}$ meson, and
\begin{align}
\phi_{\bot}^{A}(u)=6u(1-u)[1+3a_{1}^{\bot}\xiup +a_{2}^{\bot}\frac{3}{2}(5\xiup ^{2}-1)]\tag{35}
\end{align}
for $1^{1}P_{1}$ meson, where $\xiup =2u-1$, and $a_{0}^{\bot}$, $a_{1}^{\bot}$ and $a_{2}^{\bot}$ are Gegenbauer coefficients.

Matching the expression on phenomenological side with that on theoretical side, we get
\begin{equation}
-\frac{2if_{+}^{D\rightarrow S}(q^{2})p_{\mu}+i[f_{+}^{D\rightarrow S}(q^{2})+f_{-}^{D\rightarrow S}(q^{2})]q_{\mu}}{m_{D}^{2}-(p+q)^{2}}\frac{m^{2}_{D}f_{D}}{m_{c}+m_{q_{1}}}=2im_{c}p_{\mu}\int^{1}_{\Delta}du\frac{\phi(u)}{m^{2}_{c}-(q+up)^{2}},\tag{36}\\
\label{36}
\end{equation}
\begin{equation}
-\frac{[q^{2}2p_{\mu}-2q_{\mu}(p\cdot q)]}{m_{D}^{2}-(p+q)^{2}}\frac{if_{T}^{D\rightarrow S}(q^{2})}{m_{D}+m_{P}}\frac{m^{2}_{D}f_{D}}{m_{c}+m_{q_{1}}}=-[2p_{\mu}q^{2}-2q_{\mu}(p\cdot q)]\int^{1}_{\Delta}du\frac{\phi(u)}{m^{2}_{c}-(q+up)^{2}}\tag{37}\\
\end{equation}
for $D\rightarrow S$, and
\begin{align}
\{\frac{2iA^{D\rightarrow A}(q^2)}{m_{D}-m_{A}}\epsilon_{\mu\nu\alpha\beta}\varepsilon^{*\nu}(q+p)^{\alpha}p^{\beta}-(m_{D}-m_{A})V_{1}^{D\rightarrow A}(q^2)\varepsilon^*_{\mu}+\frac{V_{2}^{D\rightarrow A}(q^2)}{m_{D}-m_{A}}(\varepsilon^*\cdot{q})\notag\\
\times(2p+q)_{\mu}
+\frac{2(\varepsilon^*\cdot{q})m_{A}}{q^{2}}q_{\mu}[V_{3}^{D\rightarrow A}(q^2)-V_{0}^{D\rightarrow A}(q^2)]\}\frac{1}{m_{D}^{2}-(p+q)^{2}}\frac{m^{2}_{D}f_{D}}{m_{c}+m_{q_{1}}}\notag\\
=-2f^{A}_{\bot}\int^{1}_{\Delta}du[-p\cdot{(q+up)}\varepsilon^{*}_{\bot\mu}-(\varepsilon^*_{\bot}\cdot{q})p_{\mu}-i\epsilon_{\mu\nu\beta\tau}\varepsilon^{*\beta}
_{\bot}q^{\mu}p^{\tau}]\frac{\phi_{\bot}(u)}{m^{2}_{c}-(q+up)^{2}},\tag{38}
\end{align}
\begin{align}
\{-2\epsilon_{\mu\nu\rho\sigma}\varepsilon^{*\nu}q^{\rho}p^{\sigma}T_{1}^{D\rightarrow A}(q^2)
+i(\varepsilon^*\cdot{q})p_{\mu}[T_{2}^{D\rightarrow A}(q^2)+(\frac{q^2}{m_{D}^2-m_{A}^2}-1)T_{3}^{D\rightarrow A}(q^2)]\}\notag\\
+2i(\varepsilon^*\cdot{q})p_{\mu}[T_{2}^{D\rightarrow A}(q^2)+\frac{q^2}{m_{D}^2-m_{A}^2}T_{3}^{D\rightarrow A}(q^2)]-i\varepsilon^*_{\mu}(m_{D}^2-m_{A}^2)T_{2}^{D\rightarrow A}(q^2)\}\frac{1}{m_{D}^{2}-(p+q)^{2}}\notag\\
\times\frac{m^{2}_{D}f_{D}}{m_{c}+m_{q_{1}}}=2f^{A}_{\bot}m_{c}\int^{1}_{\Delta}du[i(q\cdot{p})\varepsilon^{*}_{\bot\mu}
-i(\varepsilon^{*}_{\bot}\cdot{q})p_{\mu}-\epsilon_{\mu\nu\rho\tau}\varepsilon^{*\rho}_{\bot}q^{\nu}p^{\tau}]\frac{\phi_{\bot}(u)}
{m^{2}_{c}-(q+up)^{2}}\tag{39}\label{39}
\end{align}
for $D\rightarrow A$, where $\Delta$ is the solution to the equation $us-m_{c}^{2}-u(1-u)m_{M}^{2}+(1-u)q^{2}=0$,
\begin{align}
\Delta&=\frac{\sqrt{(s_{0}-m_{M}^{2}-q^{2})^{2}+4(m_{c}^{2}-q^{2})m_{M}^{2}}-(s_{0}-m_{M}^{2}-q^{2})}{2m_{M}^{2}}\tag{40}  \label{40}
\end{align}
and $u\in {[0,1]}$.

In order to suppress the contribution of higher excited states and continuum, we perform Borel transformation \cite{Colangelo:2000dp} on both sides of Eps. (\ref{36})-(\ref{39}),
\begin{align}
B_{M^2}\frac{1}{m_{D}^2-(q+p)^2}&=\frac{1}{M^2}e^{-\frac{m_{D}^2}{M^2}},\notag\\
B_{M^2}\frac{1}{m_{c}^2-(q+up)^2}&=\frac{1}{uM^2}e^{-\frac{m_{c}^2+u(1-u)p^2-(1-u)q^2}{uM^2}},\tag{44}  \label{41}
\end{align}
where $M^2$ is the Borel parameter.
Up to this point, we get the analytical results of the form factors
\begin{align}
f_{+}^{D\rightarrow S}(q^2)&=-\frac{m_{c}+m_{q_{1}}}{m_{D}^{2}f_{D}}m_{c}\int^{1}_{\Delta}\frac{\phi(u)}{u}due^{FF},\notag\\
f_{-}^{D\rightarrow S}(q^2)&=\frac{m_{c}+m_{q_{1}}}{m_{D}^{2}f_{D}}m_{c}\int^{1}_{\Delta}\frac{\phi(u)}{u}due^{FF},\notag\\
f_{T}^{D\rightarrow S}(q^2)&=(m_{D}+m_{S})\frac{m_{c}+m_{q_{1}}}{m_{D}^{2}f_{D}}\int^{1}_{\Delta}\frac{\phi(u)}{u}due^{FF} \tag{42}  \label{42}
\end{align}
for $D\rightarrow S$, and
\begin{align}
A^{D\rightarrow A}(q^2)&=-\frac{m_{c}+m_{q_{1}}}{m_{D}^{2}f_{D}}(m_{D}-m_{A})f^{A}_{\bot}\int^{1}_{\Delta}\frac{\phi(u)}{u}due^{FF},\notag\\
V_{1}^{D\rightarrow A}(q^2)&=-\frac{m_{c}+m_{q_{1}}}{m_{D}^{2}f_{D}}\frac{f^{A}_{\bot}}{m_{D}-m_{A}}\int^{1}_{\Delta}\frac{\phi(u)}{u}\frac{m_{c}^2-q^2+u^2p^2}{u}due^{FF},\notag\\
V_{2}^{D\rightarrow A}(q^2)&=-\frac{m_{c}+m_{q_{1}}}{m_{D}^{2}f_{D}}(m_{D}-m_{A})f^{A}_{\bot}\int^{1}_{\Delta}\frac{\phi(u)}{u}due^{FF},\notag\\
T_{1}^{D\rightarrow A}(q^2)&=-\frac{m_{c}+m_{q_{1}}}{m_{D}^{2}f_{D}}m_{c}f^{A}_{\bot}\int^{1}_{\Delta}\frac{\phi(u)}{u}due^{FF},\notag\\
T_{2}^{D\rightarrow A}(q^2)&=-\frac{m_{c}+m_{q_{1}}}{m_{D}^{2}f_{D}}m_{c}f^{A}_{\bot}(1-\frac{q^2}{m_{D}^2-m_{A}^2})\int^{1}_{\Delta}\frac{\phi(u)}{u}due^{FF},\notag\\
T_{3}^{D\rightarrow A}(q^2)&=-\frac{m_{c}+m_{q_{1}}}{m_{D}^{2}f_{D}}m_{c}f^{A}_{\bot}\int^{1}_{\Delta}\frac{\phi(u)}{u}due^{FF},\notag
\end{align}
\begin{align}
V_{0}^{D\rightarrow A}(q^2)=&-\frac{f^{A}_{\bot}}{2m_{A}}\frac{m_{c}+m_{q_{1}}}{m_{D}^{2}f_{D}}\int^{1}_{\Delta}\frac{\phi(u)}{u}\frac{m_{c}^2-q^2+u^2p^2}{u}due^{FF}\notag\\
&+\frac{f^{A}_{\bot}}{2m_{A}}\frac{m_{c}+m_{q_{1}}}{m_{D}^{2}f_{D}}(m_{D}^{2}-m_{A}^{2})f^{A}_{\bot}\int^{1}_{\Delta}\frac{\phi(u)}{u}due^{FF}\notag\\
&+\frac{m_{c}+m_{q_{1}}}{m_{D}^{2}f_{D}}\frac{q^{2}f^{A}_{\bot}}{2m_{A}}\int^{1}_{\Delta}\frac{\phi(u)}{u}due^{FF},\notag\\
V_{3}^{D\rightarrow A}(q^2)=&-\frac{f^{A}_{\bot}}{2m_{A}}\frac{m_{c}+m_{q_{1}}}{m_{D}^{2}f_{D}}\int^{1}_{\Delta}\frac{\phi(u)}{u}\frac{m_{c}^2-q^2+u^2p^2}{u}due^{FF}\notag\\
&+\frac{f^{A}_{\bot}}{2m_{A}}\frac{m_{c}+m_{q_{1}}}{m_{D}^{2}f_{D}}(m_{D}^{2}-m_{A}^{2})f^{A}_{\bot}\int^{1}_{\Delta}\frac{\phi(u)}{u}due^{FF} \tag{43}
\end{align}
for $D\rightarrow A$, where
\begin{align}
FF&=-\frac{1}{uM^{2}}(m^{2}_{c}+u\bar{u}p^{2}-\bar{u}q^{2})+\frac{m^{2}_{D}}{M^{2}}.\tag{44}
\end{align}
Obviously, there exist some relations between form factors, which are similar to that obtained from $B\rightarrow Sl\bar{\nu}_{l}$ process \cite{Sun:2010nv,YanJun:2011rn},
\begin{align}
f_{-}^{D\rightarrow S}(q^2)&=-f_{+}^{D\rightarrow S}(q^2),\notag\\
f_{T}^{D\rightarrow S}(q^2)&=\frac{m_{D}+m_{S}}{m_{c}}f_{+}^{D\rightarrow S}(q^2),\notag\\
A^{D\rightarrow A}(q^2)&=V_{2}^{D\rightarrow A}(q^2),\notag\\
V_{0}^{D\rightarrow A}(0)&=V_{3}^{D\rightarrow A}(0),\notag\\
T_{1}^{D\rightarrow A}(q^2)&=T_{3}^{D\rightarrow A}(q^2),\notag\\
T_{2}^{D\rightarrow A}(q^2)&=(1-\frac{q^2}{m_{D}^2-m_{V}^2})T_{3}^{D\rightarrow A}(q^2).\tag{45}
\end{align}

\textbf{Branching ratios}

Using the above form factors, we can further calculate the differential decay widths of these decay processes.
For the semileptonic decay $D \rightarrow Sl\bar {\nu}_{l}$, the differential decay width can be written as \cite{Cheng:2017fkw}
\begin{align}
\frac{d\Gamma (D \rightarrow Sl\bar {\nu}_{l})}{dq^{2}}=&\frac{G_{F}^{2}|V_{cd}|^{2}}{768\pi^{3}m_{D}^{3}}\frac{(q^{2}-m_{l}^{2})^{2}}{q^{6}} \sqrt {(m_{D}^{2}+m_{S}^{2}-q^{2})^{2}-4m_{D}^{2}m_{S}^{2}}\notag\\ & \times \{(f_{+}(q^{2}))^{2}[(q^{2}+m_{S}^{2}-m_{D}^{2})^{2}(q^{2}+2m_{l}^{2})-q^{2}m_{S}^{2}(4q^{2}+2m_{l}^{2})]\notag\\
&+6f_{+}(q^{2})f_{-}(q^{2})q^{2}m_{l}^{2}(m_{D}^{2}-m_{S}^{2}-q^{2})+6(f_{-}(q^{2}))^{2}q^{4}m_{l}^{2} \}.\tag{46}  \label{46}
\end{align}
For the semileptonic decay $D \rightarrow Al\bar {\nu}_{l}$, the differential decay widths can be written as \cite{Momeni:2019uag}
\begin{align}
\frac{d\Gamma _{L}(D \rightarrow Al\bar {\nu}_{l})}{dq^{2}}=&(\frac{q^{2}-m_{l}^{2}}{q^{2}})^{2}\frac{\sqrt {\lambdaup}G_{F}^{2}|V_{cd}|^{2}}{384\pi^{3}m_{D}^{3}} \times \frac{1}{q^{2}}\{3m_{l}^{2} \lambdaup V_{0}^{2}(q^{2})+(m_{l}^{2}+2q^{2}) \notag\\ &\times |\frac{1}{2m_{A}^{2}}[(m_{D}^{2}-m_{A}^{2}-q^{2})(m_{D}-m_{A})V_{1}(q^{2})-\frac{\lambdaup}{m_{D}-m_{A}}V_{2}(q^{2})]|^{2}\} \tag{47}  \label{47}
\end{align}
and \cite{Momeni:2019uag}
\begin{align}
\frac{d\Gamma _{\pm}(D \rightarrow Al\bar {\nu}_{l})}{dq^{2}}=&(\frac{q^{2}-m_{l}^{2}}{q^{2}})^{2}\frac{\sqrt {\lambdaup}G_{F}^{2}|V_{cd}|^{2}}{384\pi^{3}m_{D}^{3}} \times \{(m_{l}^{2}+2q^{2})\lambdaup|\frac{A(q^{2})}{m_{D}-m_{A}}\mp \frac{(m_{D}-m_{A})V_{1}(q^{2})}{\lambdaup}|^{2} \},\tag{48}  \label{48}
\end{align}
where $\lambdaup=m_{D}^{4}+m_{A}^{4}+q^{4}-2m_{A}^{2}m_{D}^{2}-2q^{2}m_{A}^{2}$, $\frac{d\Gamma _{L}}{dq^{2}}$ and $\frac{d\Gamma _{\pm}}{dq^{2}}$ are the longitudinal and transverse components of the differential decay width, respectively. The total differential decay width can be written as \cite{Momeni:2019uag}
\begin{align}
\frac{d\Gamma (D \rightarrow Al\bar {\nu}_{l})}{dq^{2}}=\frac{d\Gamma _{L}(D \rightarrow Al\bar {\nu}_{l})}{dq^{2}}+\frac{d\Gamma _{\pm}(D \rightarrow Al\bar {\nu}_{l})}{dq^{2}},\tag{49}  \label{49}
\end{align}
where $G_{F}$ is Fermi coupling constant, and $|V_{cd}|$ is CKM matrix element.
\par
It should be noted that when calculating $\frac{d\Gamma}{dq^{2}}$,  due to the quark contents of scalar and axial-vector meson,
\begin{align}
|a_{0}^{0}(980)\rangle=&\frac{1}{\sqrt{2}}(|u\bar{u}\rangle-|d\bar{d}\rangle)  ,   |a_{0}^{-}(980)\rangle=|d\bar{u}\rangle,\notag\\
|a_{0}^{0}(1450)\rangle=&\frac{1}{\sqrt{2}}(|u\bar{u}\rangle-|d\bar{d}\rangle) ,   |a_{0}^{-}(1450)\rangle=|d\bar{u}\rangle,\notag\\
|a_{1}^{0}(1260)\rangle=&\frac{1}{\sqrt{2}}(|u\bar{u}\rangle-|d\bar{d}\rangle) ,   |a_{1}^{-}(1260)\rangle=|d\bar{u}\rangle,\notag\\
|b_{1}^{0}(1235)\rangle=&\frac{1}{\sqrt{2}}(|u\bar{u}\rangle-|d\bar{d}\rangle) ,   |b_{1}^{-}(1235)\rangle=|d\bar{u}\rangle,\tag{50}
\end{align}
the semileptonic decay widths for $D^{+} \rightarrow S(A)l\bar {\nu}_{l}$ corresponding to Eqs. (\ref{46})-(\ref{49}) should have an additional factor $\frac{1}{2}$.

\section{Numerical analyses and discussion}

\subsection{Choices of input parameters}

To calculate the numerical results of form factors and branching ratios, we briefly discuss the parameters involved in these semileptonic decay processes.
For the initial state, we take the masses as $m_{D^{0}}=1.865$ \rm{GeV} and $m_{D^{+}}=1.870$ \rm{GeV} \cite{Zyla:2020zbs}, decay constant as $f_{D^{0}}=f_{D^{+}}=205.4$ \rm{MeV} \cite{Amhis:2019ckw}. The quark masses are taken as $m_{u}=2.16$ \rm{MeV}, $m_{d}=4.67$ \rm{MeV} and $m_{c}=1.27$ \rm{GeV} \cite{Zyla:2020zbs}. Masses, decay constants \cite{Zyla:2020zbs} and Gegenbauer coefficients \cite{Cheng:2005nb,Yang:2008xw} for LCDAs of the final state mesons are shown in Table \uppercase\expandafter{\romannumeral2}. And the parameters \cite{Zyla:2020zbs} involved in the calculation of the branching ratios are shown in Table III.

\begin{table}[H]
	\centering
\caption{Masses, decay constants and Gegenbauer coefficients of distribution amplitudes of final state mesons at the scale $\mu=1$ \rm{GeV}.}
	\begin{tabular}{p{3cm}<{\centering}p{1.6cm}<{\centering}p{2cm}<{\centering}p{2cm}<{\centering}p{1.2cm}<{\centering}p{1.2cm}<{\centering}p{1.2cm}<{\centering}p{1.2cm}
<{\centering}p{1.2cm}<{\centering}}
		\hline
        Process &  $M$ [\rm{GeV}] & $\bar{f}_{M}$ [\rm{MeV}]& $f^{M}_{\bot}$ [\rm{MeV}]&$B_{1}$ &$B_{3}$ &$a_{0}^{\bot}$&$a_{1}^{\bot}$&$a_{2}^{\bot}$ \\
        \hline
        $ D\rightarrow a_{0}(980) $   & 0.980  & 0.365 & - & -0.93& 0.14& -& -& -\\
		
		$ D\rightarrow a_{0}(1450) $  & 1.474  & -0.280 & - & 0.89& -1.38& -& -& -\\

		$ D\rightarrow a_{1}(1260) $  & 1.230  & -  & 238 & -& -& 0& -1.04& 0\\

		$ D\rightarrow b_{1}(1235) $  & 1.230  & -  & 180 & -& -& 1& 0& 0.03 \\
		\hline
	\end{tabular}
\end{table}
\begin{table}[H]
	\centering
\caption{Parameters in branching ratios.}
	\begin{tabular}{p{2.6cm}<{\centering}p{2cm}<{\centering}p{2cm}<{\centering}p{1.5cm}<{\centering}p{3.2cm}<{\centering}p{3.2cm}<{\centering}}
		\hline
        $G_{F}$ [\rm{GeV}$^{-2}$] &$m_{e}$ $[\rm{GeV}]$ &$ m_{\mu}$ $[\rm{GeV}]$ & $|V_{cd}|$ &$\tau_{D^{+}}$ [\rm{s}] &$\tau_{D^{0}}$ [\rm{s}] \\
        \hline
        $1.1664\times 10^{-5}$ & $0.511\times 10^{-3}$ &$0.106$ &0.220 &$1.040\pm0.007\times10^{-12}$  & $4.101\pm0.015\times10^{-13}$ \\
		\hline
	\end{tabular}
\end{table}

\subsection{Dependence of form factors on threshold and Borel parameter  }

The choice of the continuum threshold and the Borel parameter is also crucial. The threshold is generally chosen as the mass square of the first excited state of $D$ meson, but this value is not universal and has to be determined individually according to the sum rule of different processes. The choice of the Borel parameter should satisfy:
\par
\textbf{(1)} the contributions from the continuum and higher excited states are less than 30\%;
\par
\textbf{(2)} the dependence of form factors on the Borel parameter is weak. \\
Based on the above conditions, the corresponding thresholds and the Borel windows for different semileptonic decay processes are ascertained, as is shown in Fig. 1-4.

For the convenience of discussion in the following, we take the central value of threshold as $s_{0}=5.18$ \rm{GeV}$^{2}$ for $ D\rightarrow a_{0}(980) $, $ D\rightarrow a_{0}(1450)$, $ D\rightarrow a_{1}(1260) $ and $D\rightarrow b_{1}(1235)$  respectively.
\begin{figure}[H]
\centering
\includegraphics[width=7cm,height=6cm]{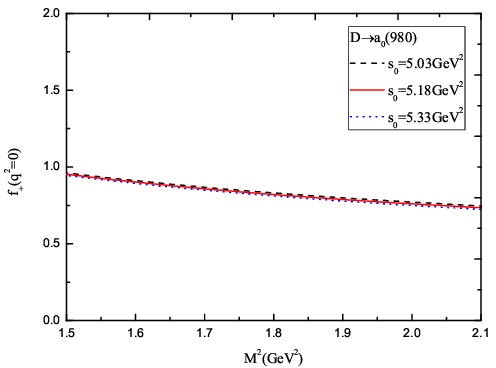}
\includegraphics[width=7cm,height=6cm]{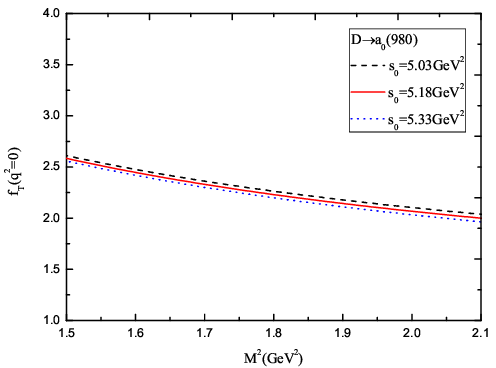}
\caption{Dependence of $ D\rightarrow a_{0}(980) $ transition form factors and penguin form factors on the Borel parameter $M^{2}$ at $q^{2}=0$ \rm{GeV}$^{2}$. The dash, solid and dot lines correspond to thresholds $s_{0}=5.03$, $5.18$ and $5.33$ \rm{GeV}$^{2}$ respectively.}
\end{figure}
\begin{figure}[H]
\centering
\includegraphics[width=7cm,height=6cm]{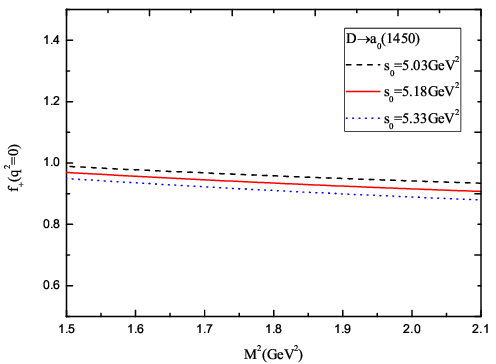}
\includegraphics[width=7cm,height=6cm]{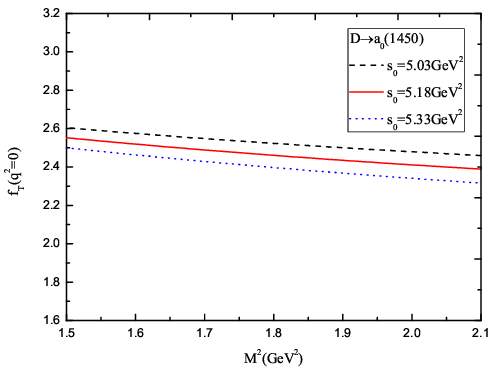}
\caption{Dependence of $ D\rightarrow a_{0}(1450)$ transition form factors and penguin form factors on the Borel parameter $M^{2}$ at $q^{2}=0$ \rm{GeV}$^{2}$. The dash, solid and dot lines correspond to thresholds $s_{0}=5.03$, $5.18$ and $5.33$ \rm{GeV}$^{2}$ respectively.}
\end{figure}
\begin{figure}[H]
\centering
\includegraphics[width=7cm,height=6cm]{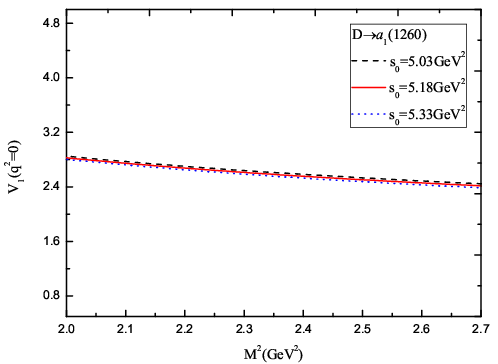}
\includegraphics[width=7cm,height=6cm]{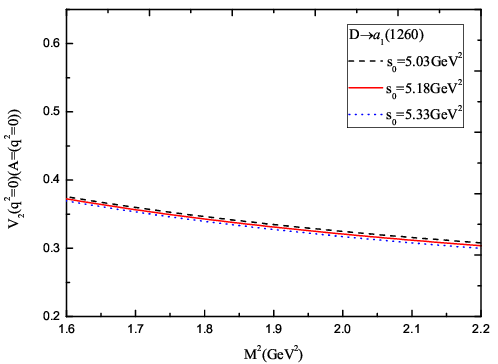}
\includegraphics[width=7cm,height=6cm]{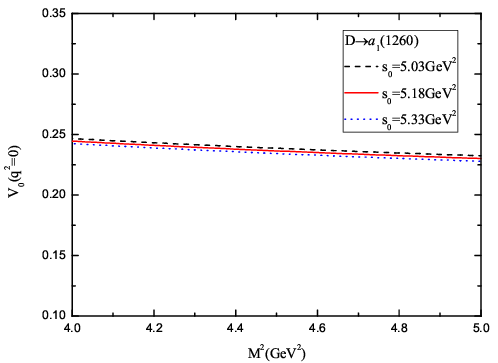}
\includegraphics[width=7cm,height=6cm]{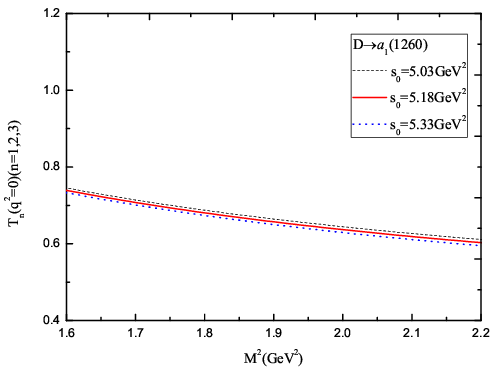}
\caption{Dependence of $ D\rightarrow a_{1}(1260) $ transition form factors and penguin form factors on the Borel parameter $M^{2}$ at the transfer momentum $q^{2}=0$ \rm{GeV}$^{2}$. The dash, solid and dot lines correspond to thresholds $s_{0}=5.03$, $5.18$ and $5.33$ \rm{GeV}$^{2}$ respectively. }
\end{figure}
\begin{figure}[H]
\centering
\includegraphics[width=7cm,height=6cm]{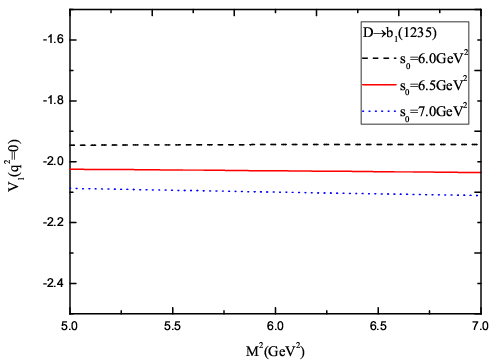}
\includegraphics[width=7cm,height=6cm]{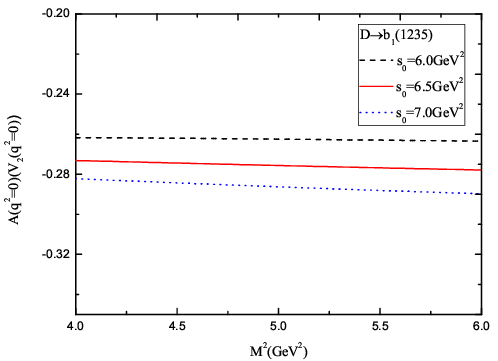}
\includegraphics[width=7cm,height=6cm]{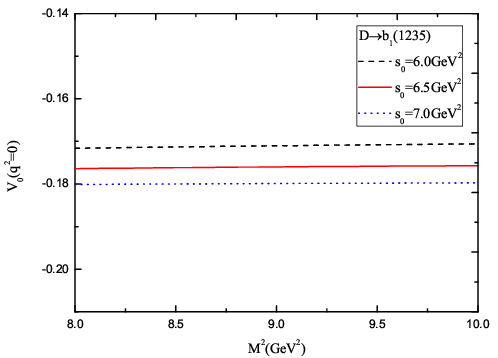}
\includegraphics[width=7cm,height=6cm]{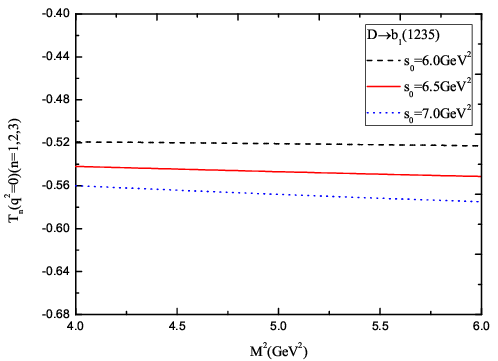}
\caption{Dependence of $ D\rightarrow b_{1}(1235)$ transition form factors and penguin form factors on the Borel parameter $M^{2}$ at the transfer momentum $q^{2}=0$ \rm{GeV}$^{2}$. The dash, solid and dot lines correspond to thresholds $s_{0}=5.03$, $5.18$ and $5.33$ \rm{GeV}$^{2}$ respectively. }
\end{figure}

\subsection{Results and discussion}
\textbf{The form factors at $q^{2}= 0$ \rm{GeV}$^{2}$}

Based on the above parameters, we get the numerical results of form factors in case of $q^2=0$ \rm{GeV}$^{2}$, shown in Table IV and Table V .

\begin{table}[H]
	\centering
    \caption{Comparison of numerical results of form factors for $D\rightarrow S$ with that of Covariant Confined Quark Model(CCQM) and Covariant Light-Front Quark Model(CLFQM).}
	\begin{tabular}{|c|c|c|c|c|}
		\hline
       Process& Method & $f_{+}(0)$ & $f_{-}(0)$  & $f_{T}(0)$\\
        \hline
		\multirow{3}*{$ D^{0}\rightarrow a_{0}^{-}(980) $ } & this work & $0.85 ^{+0.10}_{-0.11}$& $-0.85 ^{+0.10}_{-0.11}$& $2.29 ^{+0.29}_{-0.30}$ \\
		\cline{2-5}
		~ & LCSR\cite{Cheng:2017fkw} & $1.75 ^{+0.26}_{-0.27}$  & $0.31\pm 0.13$ &- \\
        \cline{2-5}
        ~ & CCQM\cite{Soni:2020sgn} & $0.55\pm 0.02$ & $0.03\pm 0.01$ & - \\
        \hline
		\multirow{3}*{$ D^{+}\rightarrow a_{0}^{0}(980) $ } & this work & $0.85 ^{+0.10}_{-0.11}$& $-0.85 ^{+0.10}_{-0.11}$& $2.29 ^{+0.29}_{-0.30}$ \\
		\cline{2-5}
		~ & LCSR\cite{Cheng:2017fkw} & $1.76 \pm0.26$  & $0.31\pm 0.13$ &- \\
        \cline{2-5}
        ~ & CCQM\cite{Soni:2020sgn} & $0.55\pm 0.02$ & $0.03\pm 0.01$ & - \\
		\hline
         \multirow{2}*{$ D^{0}\rightarrow a_{0}^{-}(1450) $ }& this work & $0.94 ^{+0.02}_{-0.03}$ &$-0.94 ^{+0.02}_{-0.03}$ &$2.48 ^{+0.07}_{-0.08}$\\
         \cline{2-5}
        ~ & CLFQM\cite{Verma:2011yw} & $0.51^{-0.01+0.01}_{+0.01-0.02}$ & - & - \\
        \hline
         \multirow{2}*{$ D^{+}\rightarrow a_{0}^{0}(1450) $ }& this work & $0.94 ^{+0.02}_{-0.03}$ &$-0.94 ^{+0.02}_{-0.03}$ &$2.48 ^{+0.07}_{-0.08}$\\
         \cline{2-5}
        ~ & CLFQM\cite{Verma:2011yw} & $0.51^{-0.01+0.01}_{+0.01-0.02}$ & - & - \\
		\hline
	\end{tabular}
\end{table}

For $ D\rightarrow a_{0}(980) $, our results for $f_{+}(0)$ are $51 \% $ smaller than those of Ref. \cite{Cheng:2017fkw}, mainly due to the different input parameter $\bar{f}_{s}$ and the inclusion of twist-3 LCDAs in Ref. \cite{Cheng:2017fkw}.

\begin{table}[H]
	\centering
    \caption{Comparison of numerical results of form factors for $D\rightarrow A$ with that of LCSR, 3-point QCD sum rules (3QR) and CLFQM (n=1,2,3). }
	\begin{tabular}{|c|c|c|c|c|c|c|}
		\hline
       Process& Method & $A(0)$ &$ V_{1}(0)$ &$V_{2}(0)$& $V_{0}(0)$  & $T_{n}(0)$\\
       \hline
		\multirow{4}*{$ D^{0} \rightarrow a_{1}^{-}(1260) $} & this work & $0.34^{+0.03}_{-0.04}$ & $2.63^{+0.20}_{-0.21}$ & $0.34^{+0.03}_{-0.04}$ & $0.24^{+0.00}_{-0.01}$ & $0.67^{+0.07}_{-0.07}$\\
		\cline{2-7}
		~ & LCSR \cite{Momeni:2019uag} & $0.07\pm 0.05$ &$0.37\pm 0.01 $ & $-0.03\pm 0.02$  &$0.15\pm 0.05$  & -\\
       \cline{2-7}
		~ & 3SR \cite{Zuo:2016msr} & $0.09$ & $0.77$ & $-0.01$ &$0.19$  & -\\
       \cline{2-7}
        ~ & CLFQM \cite{Verma:2011yw} & $0.19^{-0.01+0.00}_{+0.01-0.00}$ & $1.51^{-0.04+0.00}_{+0.04-0.01}$ & $0.05^{-0.01+0.00}_{+0.00-0.00}$ & $0.32^{-0.00-0.00}_{+0.00-0.00}$ & -\\
        \hline
		\multirow{4}*{$ D^{+} \rightarrow a_{1}^{0}(1260) $} & this work & $0.34^{+0.03}_{-0.04}$ & $2.63^{+0.20}_{-0.21}$ & $0.34^{+0.03}_{-0.04}$ & $0.24^{+0.00}_{-0.01}$ & $0.67^{+0.07}_{-0.07}$\\
		\cline{2-7}
		~ & LCSR \cite{Momeni:2019uag} & $0.04\pm 0.04$ & $0.26\pm 0.08$ & $-0.02\pm 0.01$ &$0.10\pm 0.03$  & -\\
        \cline{2-7}
		~ & 3SR \cite{Zuo:2016msr} & $0.08$ & $0.54$ & $-0.00$ &$0.12$  & -\\
        \cline{2-7}
       ~ & CLFQM \cite{Verma:2011yw} & $0.19^{-0.01+0.00}_{+0.01-0.00}$ & $1.51^{-0.04+0.00}_{+0.04-0.01}$ & $0.05^{-0.01+0.00}_{+0.00-0.00}$ & $0.32^{-0.00-0.00}_{+0.00-0.00}$ & -\\
        \hline
		\multirow{3}*{$ D^{0} \rightarrow b_{1}^{-}(1235) $} & this work & $-0.24^{+0.01}_{-0.00}$ & $-1.78^{+0.03}_{-0.02}$ & $-0.24^{+0.01}_{-0.00}$ & $-0.16^{+0.00}_{-0.00}$ & $-0.47^{+0.01}_{-0.01}$\\
		\cline{2-7}
		~ & LCSR \cite{Momeni:2019uag} & $-0.41$ & $-0.22$ & 0.21 &$-0.32$  & -\\
       \cline{2-7}
        ~ & CLFQM \cite{Verma:2011yw} & $0.12^{+0.00+0.00}_{-0.00-0.02}$ & $1.39^{+0.02+0.03}_{-0.02-0.04}$ & $-0.10^{+0.02-0.01}_{-0.02+0.01}$ & $0.50^{-0.01+0.02}_{+0.01-0.02}$ & -\\
        \hline
		\multirow{3}*{$ D^{+} \rightarrow b_{1}^{0}(1235) $} & this work & $-0.24^{+0.01}_{-0.00}$ & $-1.78^{+0.03}_{-0.02}$ & $-0.24^{+0.01}_{-0.00}$ & $-0.16^{+0.00}_{-0.00}$ & $-0.47^{+0.01}_{-0.01}$\\
		\cline{2-7}
		~ & LCSR \cite{Momeni:2019uag} & $-0.28$ & $-0.16$ & 0.15 &$-0.23$  & -\\
       \cline{2-7}
        ~ & CLFQM \cite{Verma:2011yw} & $0.12^{+0.00+0.00}_{-0.00-0.02}$  & $1.39^{+0.02+0.03}_{-0.02-0.04}$ & $-0.10^{+0.02-0.01}_{-0.02+0.01}$ &$0.50^{-0.01+0.02}_{+0.01-0.02}$ & -\\
		\hline
    \end{tabular}
\end{table}

In Table V we give the form factors for $D\rightarrow a_{1}(1260)$ and $D\rightarrow b_{1}(1235)$, which is larger compared to the results of Ref. \cite{Momeni:2019uag}, since contribution from the high twist LCDAs is included.

\textbf{The form factors at $q^{2}\neq 0$ \rm{GeV}$^{2}$}

Fig. 3 shows the dependence of the form factors on $ m_{l}^{2}\leq q^{2}\leq(m_{D}-m_{S})^{2} $  for $ D\rightarrow S $. For $ D \rightarrow a_{0}(980) $, we take the threshold as $ s_{0}=5.18$ \rm{GeV}$^{2}$ and the Borel parameter as $M^{2}=1.72$ \rm{GeV}$^{2}$. For $ D\rightarrow a_{0}(1450) $, we take the threshold as $ s_{0}=5.18$ \rm{GeV}$^{2} $ and the Borel parameter as $ M^{2}=1.74$ \rm{GeV}$^{2}$.

\begin{figure}[H]
\centering
\includegraphics[width=7cm,height=6cm]{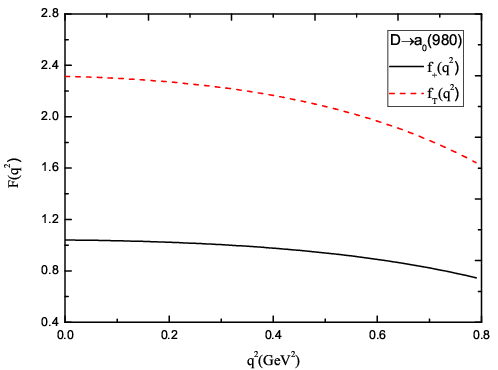}
\includegraphics[width=7cm,height=6cm]{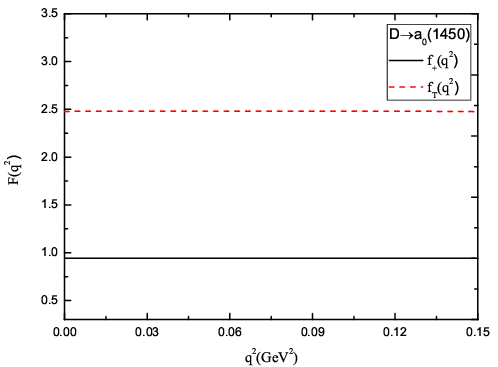}
\caption{Dependence of the form factors on the transfer momentum $q^{2}$ for $ D\rightarrow a_{0}(980) $ and $ D\rightarrow a_{0}(1450)$ decay. The solid and dash lines depict the form factors $f_{+}(q^{2})$, $f_{T}(q^{2})$ , respectively.}
\end{figure}

Fig. 4 shows the dependence of the form factors on $m_{l}^{2}\leq q^{2}\leq(m_{D}-m_{A})^{2}$ for $D\rightarrow A$. For $ D \rightarrow a_{1}(1260)$, we take the threshold $s_{0}=5.18$ \rm{GeV}$^{2}$ and the Borel parameter $M^{2}=4$ \rm{GeV}$^{2}$, and for $ D \rightarrow b_{1}(1235)$, we take the threshold $s_{0}=5.18$ \rm{GeV}$^{2}$ and the Borel parameter $M^{2}=6$ \rm{GeV}$^{2}$.
\begin{figure}[H]
\centering
\includegraphics[width=7cm,height=6cm]{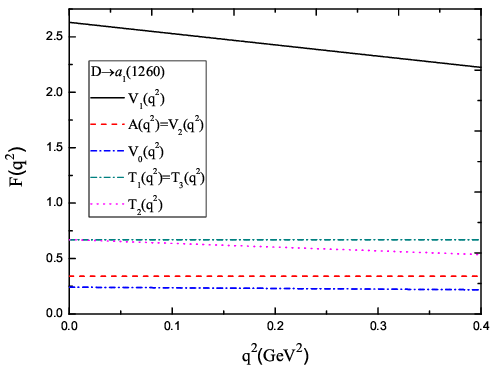}
\includegraphics[width=7cm,height=6cm]{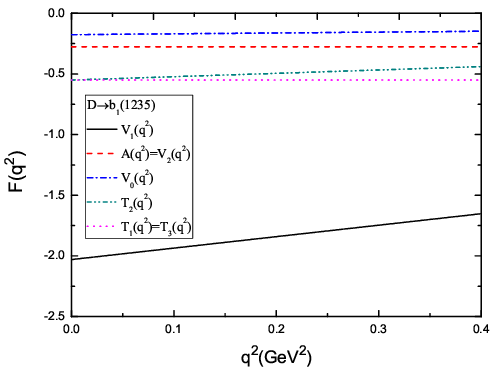}
\caption{Dependence of the form factors decay on the transfer momentum $q^{2}$ for $ D\rightarrow a_{1}(1260) $ and $ D\rightarrow b_{1}(1235) $ . The solid, dash, dash-dot, dash-dot-dot, and dot lines depict the form factors $V_{1}(q^{2})$, $A(q^{2})(V_{2}(q^{2}))$, $V_{0}(q^{2})$, $T_{1}(q^{2})(T_{3}(q^{2}))$, $T_{2}(q^{2})$, respectively.}
\end{figure}

\textbf{Branching ratios}

Fig. 5 and Fig. 6 show the dependence of the differential decay widths on the transfer momentum $q^{2}$ for $D\rightarrow Sl\bar{\nu}_{l}$ and $D\rightarrow Al\bar{\nu}_{l}$, respectively.

\begin{figure}[H]
\centering
\includegraphics[width=7cm,height=6cm]{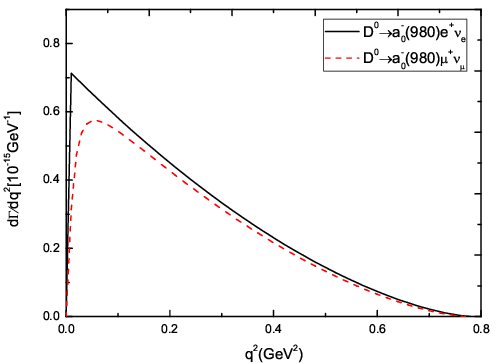}
\includegraphics[width=7cm,height=6cm]{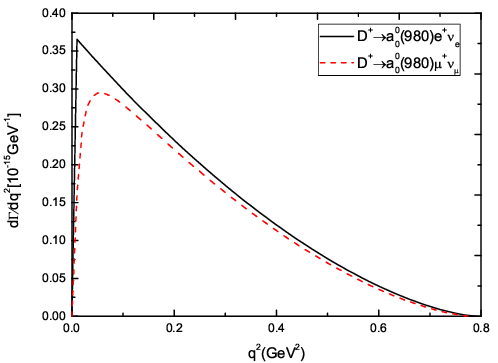}
\includegraphics[width=7cm,height=6cm]{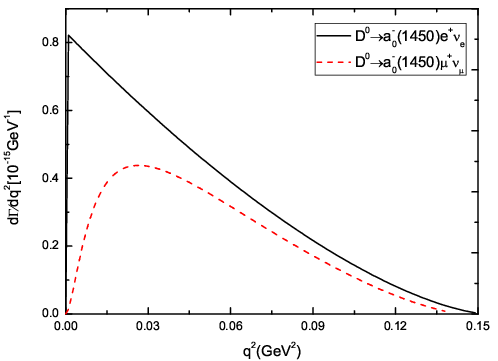}
\includegraphics[width=7cm,height=6cm]{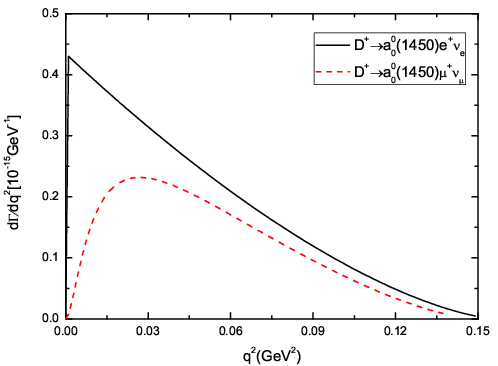}
\caption{$ D\rightarrow a_{0}(980)l\bar{\nu}_{l} $ differential decay widths at $m_{e}^{2}\leq q^{2}\leq (m_{D}-m_{a_{0}(980)})^{2}$ , and $ D\rightarrow a_{0}(1450)l\bar{\nu}_{l} $  differential decay widths at $m_{e}^{2}\leq q^{2}\leq (m_{D}-m_{a_{0}(1450)})^{2}$. }
\end{figure}
\begin{figure}[H]
\centering
\includegraphics[width=7cm,height=6cm]{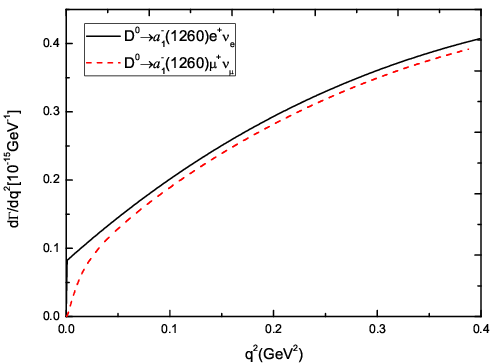}
\includegraphics[width=7cm,height=6cm]{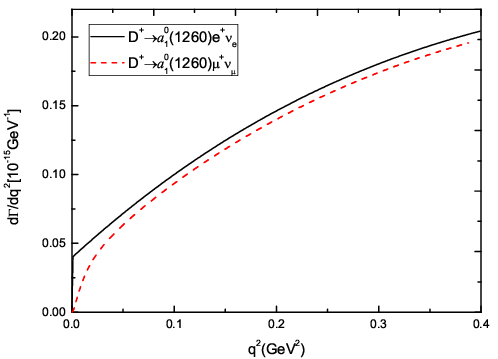}
\includegraphics[width=7cm,height=6cm]{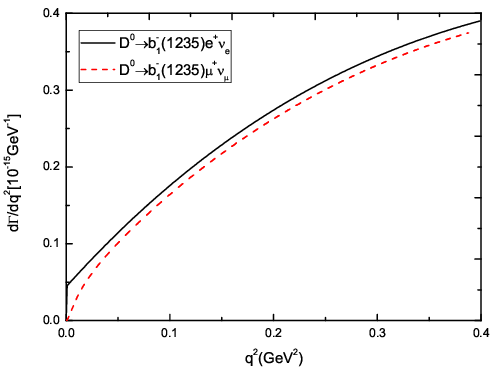}
\includegraphics[width=7cm,height=6cm]{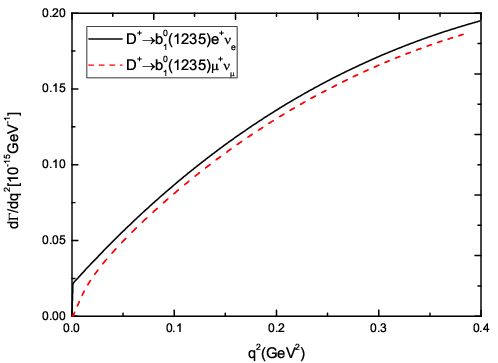}
\caption{$ D\rightarrow a_{1}(1260)l\bar{\nu}_{l} $  differential decay widths at $m_{e}^{2}\leq q^{2}\leq (m_{D}-m_{a_{1}(1260)})^{2}$, and $ D\rightarrow b_{1}(1235)l\bar{\nu}_{l} $ differential decay widths at $m_{e}^{2}\leq q^{2}\leq (m_{D}-m_{b_{1}(1235)})^{2}$ .}
\end{figure}

For differential decay widths, integrating over a range of $m_{e}^{2}\leq q^{2}\leq (m_{D}-m_{M})^{2}$, we obtain the decay widths and hence the decay branching ratios, the numerical results of which are shown in Tables VI and VII.

\begin{table}[H]
	\centering
    \caption{Comparison of numerical results of branching ratios for $D\rightarrow Sl\bar{\nu}_{l}$ with other theoretical methods and experiments. }
	\begin{tabular}{|c|c|c|c|}
		\hline
       Process& Method & Branching ratios& $\mathcal {B} (D^{0(+)}\rightarrow Sl\bar{\nu}_{e})\times \mathcal {B} (S\rightarrow \etaup\pi)$\\
       \hline
		\multirow{4}*{$ D^{0}\rightarrow a_{0}^{-}(980)e^{+}\nu_{e} $ } & this work & $1.36\times 10^{-4}$&$1.15\times 10^{-4}$\\
        \cline{2-4}
        ~ & CCQM \cite{Soni:2020sgn} &  $1.68\pm 0.15\times  10^{-4}$&$1.84\times 10^{-4}$\\
        \cline{2-4}
        ~ & LCSR \cite{Cheng:2017fkw} &  $4.08^{+1.37}_{-1.22}\times 10^{-4}$&$3.45\times 10^{-4}$\\
        \cline{2-4}
        ~ &BES \uppercase\expandafter{\romannumeral3} \cite{Ablikim:2018ffp} & - &$1.33^{+0.33}_{-0.29}\times 10^{-4} $\\
       \hline
		\multirow{2}*{$ D^{0}\rightarrow a_{0}^{-}(980)\mu^{+}\nu_{\mu} $ } & this work & $1.21\times 10^{-4}$&-\\
        \cline{2-4}
        ~ & CCQM \cite{Soni:2020sgn} &  $1.63\pm 0.14\times  10^{-4}$&-\\
        \hline
		\multirow{4}*{$ D^{+}\rightarrow a_{0}^{0}(980)e^{+}\nu_{e} $ } & this work & $1.79\times 10^{-4}$&$1.51\times 10^{-4}$\\
        \cline{2-4}
        ~ & CCQM \cite{Soni:2020sgn} &  $2.18\pm 0.38\times  10^{-4}$&$1.84\times 10^{-4}$\\
        \cline{2-4}
        ~ & LCSR \cite{Cheng:2017fkw} &  $5.40^{+1.78}_{-1.59}\times 10^{-4}$&$4.56\times 10^{-4}$\\
        \cline{2-4}
        ~ & BES \uppercase\expandafter{\romannumeral3} \cite{Ablikim:2018ffp} & -&$1.66^{+0.81}_{-0.66}\times 10^{-4} $\\
        \hline
		\multirow{2}*{$ D^{+}\rightarrow a_{0}^{0}(980)\mu^{+}\nu_{\mu} $ } & this work & $1.59\times 10^{-4}$&-\\
        \cline{2-4}
        ~ & CCQM \cite{Soni:2020sgn} &  $2.12\pm 0.37\times  10^{-4}$&-\\
		\hline
         \multirow{1}*{$ D^{0}\rightarrow a_{0}^{-}(1450)e^{+}\nu_{e} $ }& this work & $3.14\times  10^{-6}$&$0.29\times  10^{-6}$\\
        \hline
         \multirow{1}*{$ D^{0}\rightarrow a_{0}^{-}(1450)\mu^{+}\nu_{\mu} $ }& this work & $2.01\times  10^{-6}$&-\\
        \hline
         \multirow{2}*{$ D^{+}\rightarrow a_{0}^{0}(1450)e^{+}\nu_{e} $ }& this work & $4.28\times  10^{-6}$&$0.40\times  10^{-6}$\\
         \cline{2-4}
        ~ & CLFQM \cite{Cheng:2017pcq} & $5.4\pm0.5\times  10^{-6} $&-\\
        \hline
         \multirow{2}*{$ D^{+}\rightarrow a_{0}^{0}(1450)\mu^{+}\nu_{\mu} $ }& this work & $2.76\times  10^{-6}$&-\\
         \cline{2-4}
        ~ & CLFQM \cite{Cheng:2017pcq} & $3.8\pm0.3\times  10^{-6} $&-\\
		\hline
	\end{tabular}\\
\end{table}

In Table VI, for $ D^{0(+)}\rightarrow a_{0}^{-(0)}(980)e^{+}\nu_{e} $ , it is easy to see that our results are well consistent with those of BES \uppercase\expandafter{\romannumeral3}  \cite{Ablikim:2018ffp} within the error, where \cite{Cheng:2013fba}
\begin{align}
\mathcal {B} (a_{0}(980)\rightarrow \etaup\pi)&= 0.845 \pm 0.017.\tag{51}  \label{51}
\end{align}
The difference between our results and the results of LCSR \cite{Cheng:2017fkw} methods is mainly caused by form factors. It is easy to see in Eq (\ref{46}) that $(f_{+}(q^{2}))^{2}$ makes the main contribution in the differential decay width, with $(f_{+}(q^{2})f_{-}(q^{2}))$ and $f_{-}(q^{2})$ being suppressed by $m_{l}^{2}$. We also calculate the ratio of the partial width

\begin{align}
\frac{\Gamma (D^{0}\rightarrow a_{0}^{-0}(980)e^{+}\nu_{e})}{\Gamma (D^{+}\rightarrow a_{0}^{0}(980)e^{+}\nu_{e})}=2.17,\tag{52} \label{52}
\end{align}
which is consistent with the prediction $2.03\pm0.95\pm0.06$ by BES \uppercase\expandafter{\romannumeral3} \cite{Ablikim:2018ffp} within the error.

For $D\rightarrow a_{0}(1450)l\bar{\nu}_{l} $, we predict the results of the branching ratios at $\mathcal {B} (a_{0}(1450)\rightarrow \etaup\pi)=0.093$ \cite{Zyla:2020zbs} and also calculate the partial width ratios,
\begin{align}
\frac{\Gamma (D^{0}\rightarrow a_{0}^{-0}(1450)e^{+}\nu_{e})}{\Gamma (D^{+}\rightarrow a_{0}^{0}(1450)e^{+}\nu_{e})}=2.03,\tag{53} \label{53}
\end{align}
expecting them to be tested experimentally in the future.

\begin{table}[H]
	\centering
    \caption{Comparison of numerical results of branching ratios of $D\rightarrow Al\bar{\nu}_{l}$ with the results of other theoretical methods and experiments. }
	\begin{tabular}{|c|c|c|c|}
		\hline
        Process & Method & Branching ratios &$\mathcal {B}(D\rightarrow b_{1}(1235)l\bar{\nu}_{l})\times \mathcal {B}(b_{1}(1235)\rightarrow \omegaup\pi)$\\
        \hline
		\multirow{1}*{$ D^{0}\rightarrow a_{1}^{-}(1260)e^{+}\nu_{e} $} & this work & $6.90\times 10^{-5}$ & -\\
		\hline
        \multirow{1}*{$ D^{0}\rightarrow a_{1}^{-}(1260)\mu^{+}\nu_{\mu} $ } & this work & $6.27\times 10^{-5}$ & -\\
		\hline
        \multirow{1}*{$ D^{+}\rightarrow a_{1}^{0}(1260)e^{+}\nu_{e} $ } & this work & $9.38\times 10^{-5}$ & -\\
		\hline
        \multirow{1}*{$D^{+}\rightarrow a_{1}^{0}(1260)\mu^{+}\nu_{\mu} $ } & this work & $8.52\times 10^{-5}$ & -\\
        \hline
		\multirow{2}*{$ D^{0}\rightarrow b_{1}^{-}(1235)e^{+}\nu_{e} $} & this work & $4.85\times 10^{-5}$ & $4.85\times 10^{-5}$\\
        \cline{2-4}
        ~ & BES \uppercase\expandafter{\romannumeral3} \cite{Ablikim:2020agq} & - &$<1.12\times 10^{-4}$\\
		\hline
        \multirow{1}*{$ D^{0}\rightarrow b_{1}^{-}(1235)\mu^{+}\nu_{\mu} $ } & this work & $4.40\times 10^{-5}$ & -\\
		\hline
        \multirow{3}*{$ D^{+}\rightarrow b_{1}^{0}(1235)e^{+}\nu_{e} $ } & this work & $6.58\times 10^{-5}$ &$6.58\times 10^{-5}$\\
        \cline{2-4}
        ~ & CLFQM \cite{Cheng:2017pcq} &  $7.4\pm0.7\times  10^{-5}$ &$7.4\pm0.7\times  10^{-5}$\\
         \cline{2-4}
        ~ & BES \uppercase\expandafter{\romannumeral3} \cite{Ablikim:2020agq} & - &$<1.75\times 10^{-4}$\\
		\hline
        \multirow{2}*{$D^{+}\rightarrow b_{1}^{0}(1235)\mu^{+}\nu_{\mu} $ } & this work & $6.00\times 10^{-5}$ & -\\
        \cline{2-4}
        ~ & CLFQM \cite{Cheng:2017pcq} &  $6.4\pm0.6\times  10^{-5}$ & -\\
		\hline
	\end{tabular}
\end{table}

For $D\rightarrow a_{1}(1260)l\bar{\nu}_{l}$ and $D^{0(+)}\rightarrow b^{-(0)}_{1}(1235)le^{+}\nu_{e}$, BES \uppercase\expandafter{\romannumeral3} gave an upper limit for $\mathcal {B}(D^{0(+)}\rightarrow b_{1}^{-(0)}(1235)e^{+}\nu_{e})\times \mathcal {B}(b_{1}^{-(0)}(1235)\rightarrow \omegaup\pi^{-(0)})$ \cite{Ablikim:2020agq}. When taking $\mathcal {B}(b_{1}(1235)\rightarrow \omegaup\pi$)=1 \cite{Zyla:2020zbs}, our results for $D^{0(+)}\rightarrow b^{-(0)}_{1}(1235)le^{+}\nu_{e}$ are consistent with the experiment within the error, while that for $D\rightarrow a_{1}(1260)l\bar{\nu}_{l}$ are expected to be tested by experiment in the future.

\section{Summary}

We systematically study the semileptonic decay process of $ D\rightarrow S, A l \bar{\nu_{l}}$ by LCSR with chiral currents. For $ D\rightarrow S l \bar{\nu_{l}}$  we take $a_{0}(980)$ and  $a_{0}(1450)$ meson as $q\bar{q}$ states, expecting our results to be helpful in determining the hadron structures of $a_{0}(980)$ and  $a_{0}(1450)$. For $ D\rightarrow A l \bar{\nu_{l}}$, we study $a_{1}(1260)( 1^{3}p^{1})$ and $b_{1}(1235)( 1^{1}p^{1})$ mesons.
Due to the chiral currents, our methods eliminate the contributions of the higher twist LCDAs, which avoid the uncertainty from the high twist LCDAs. Simple relations between form factors for $D\rightarrow S, A l \bar{\nu_{l}}$ are also obtained, which are similar to that obtained from the $ B\rightarrow S, A l \bar{\nu_{l}}$ process \cite{Sun:2010nv,YanJun:2011rn}.
\par
We also systematically analyze the dependence of form factors and differential decay widths on the transfer momentum $q^{2}$. Applying the results of form factors and differential decay widths, we present the branching ratios of these semileptonic decay processes. Our numerical results for $ D\rightarrow a_{0}(980), b_{1}(1235)l \bar{\nu_{l}} $ are in good agreement with experiments and that for the $ D\rightarrow a_{0}(1450), a_{1}(1260)l \bar{\nu_{l}}$ processes are expected to be tested experimentally in the future.

\section*{ACKNOWLEDGEMENTS}
\par
Y.J. Sun would like to thank Y.M. Wang for helpful discussions. This work was supported in part by Natural Science Foundation of China under Grant Nos.11365018, 11375240 and 11565023.



\end{document}